\documentclass[twocolumn,prl]{revtex4} 
\usepackage{epsfig}
\begin{document} 

\title{Evidence for Quantized Displacement in Macroscopic Nanomechanical Oscillators}  
\author{Alexei Gaidarzhy$^{1,2}$, Guiti Zolfagharkhani$^1$, Robert L. Badzey$^1$, and Pritiraj Mohanty$^1$} 
\affiliation{$^1$Department of Physics, Boston University, 590 Commonwealth Avenue, Boston, MA 02215}    
\affiliation{$^2$Aerospace and Mechanical Engineering, Boston University, 110 Cummington Street, Boston, MA 02215}

\begin{abstract} 
We report the observation of discrete displacement of nanomechanical oscillators with gigahertz-range resonance frequencies at millikelvin temperatures. The oscillators are nanomachined single-crystal structures of silicon, designed to provide two distinct sets of coupled elements with very low and very high frequencies. With this novel design, femtometer-level displacement of the frequency-determining element is amplified into collective motion of the entire micron-sized structure. The observed discrete response possibly results from energy quantization at the onset of the quantum regime in these macroscopic nanomechanical oscillators.
\vskip 0.2 in  
\end{abstract}  

\maketitle 

\makeatletter
\global\@specialpagefalse
\def\@oddhead{\hskip 5.0in {Phys. Rev. Lett. {\bf 94}, 030402 (2005)}}
\let\@evenhead\@oddhead
\makeatother  

The quantum mechanical harmonic oscillator is a fundamental example of textbook quantum mechanics \cite{dirac}.  Its direct experimental realization in truly macroscopic mechanical systems is of interest to a wide range of fields \cite{leggett1,leggett2,leggett3}, which include quantum measurement \cite{zurek1,zurek2,caves,bocko}, quantum computation \cite{nakamura}, atomic and quantum optics \cite{brune}, condensed matter physics \cite{clarke,russo,friedman} and gravitational wave detection \cite{caves,bocko}.  However, despite intense experimental efforts \cite{leggett3,blencowe,cho}, particularly with nanomechanical structures \cite{blencowe-review,cross,zwerger}, signatures of quantum behavior such as quantum transitions in a macroscopic mechanical oscillator are yet to be observed \cite{caves,bocko}. 

The essential problem in achieving quantized behavior in mechanical structures \cite{braginsky} has been the access to the quantum regime. Two characteristic time scales, decoherence time and dissipation time, define quantum-to-classical crossover. Although decoherence imposes a much stricter condition, a necessary requirement for observing quantum behavior is given by dissipation or energy relaxation ($1/Q$, inverse quality factor): for a system with $Q = 1$, the quantum of oscillator energy $hf$ is larger than or comparable to $k_BT$.   Realization of this criterion requires both millikelvin temperatures and gigahertz range frequencies.  For example, a nanomechanical beam with a resonance frequency of 1 GHz will enter the quantum regime at  $T = 48\mbox{ mK}$. For a doubly-clamped beam, the fundamental frequency scales as $\sqrt{E/\rho}(t/L^2)$, where $E$ is the Young's modulus, $\rho$ is the mass density, $t$ and $L$ are thickness and length respectively.  In typical materials like silicon, all dimensions must be in the sub-micron range to achieve gigahertz resonance frequencies. However, if structure dimensions are reduced to increase the resonance frequency, it naturally increases the spring constant $k$. As the structure becomes stiffer, the displacement on resonance, $x = FQ/k$, decreases for a given amplitude of force $F$.  For a gigahertz-range beam, the typical displacement is on the order of a femtometer.  Detecting femtometer displacements is further impeded because the quality factor is known to decrease with decreasing system dimensions \cite{mohanty,carr,ahn}.

Propelled by the recent advances in nanomechanics, there have been numerous attempts to approach the quantum regime in nanomechanical oscillators \cite{blencowe,cho} with low thermal occupation number $N_{th} \equiv k_BT/hf$. The central thrust of this effort has been the development of ultra-sensitive displacement detection techniques. These include the coupling of the nanomechanical beam to a single-electron transistor (SET) sensor \cite{devoret,blencowe2,hastings}, a cooper-pair box device \cite{blencowe3}, SQUID sensor \cite{paik}, and piezoelectric sensor \cite{cleland1} as well as optical interferometric techniques \cite{rugar}.  Recently, in a 116-MHz beam oscillator measured down to a cryostat temperature of 30 mK, Knobel et al \cite{cleland-nature} have reported displacement sensitivity one hundred times the standard quantum limit, which is the limiting factor despite the very low occupation numbers they achieve.  At a lower oscillator frequency of 20 MHz measured down to the oscillator temperature of 56 mK, LaHaye et al \cite{schwab} have achieved greater resolution, 4.3 times the standard quantum limit, at the cost of larger $N_{th}=58$.  Our approach is inherently different in that instead of attempting to further improve the detection sensitivity, we focus on a novel design of the nanomechanical structures, which display gigahertz range frequencies with a corresponding displacement in the picometer range. 

In this Letter, we report the first observation of discrete displacement of possible quantum origin in a set of nanomechanical oscillators, which resonate at a frequency as high as 1.49 GHz. At a measured temperature of 110 mK, which corresponds to a thermal occupation of $N_{th} \rightarrow 1$, the oscillators demonstrate transitions between two discrete positions, with consistent amplitude in both magnetic field and time domain sweeps. We argue that the wave functions of the two low-lying energy levels of the oscillator at $N_{th} \rightarrow 1$ result in the observed quantized displacement. Furthermore, the nanomechanical structure truly represents a macroscopic quantum system as the quantized displacement involves roughly 50 billion silicon atoms.   

Our device is an antenna-like structure designed to have coupled but distinct components, pictured in Fig.~1.  Two rows of small paddles on both sides of a doubly-clamped beam serve as frequency-determining elements, resonating at the same gigahertz natural frequency.  When all paddles vibrate in phase, shown in Fig.~1c, the central beam couples to the paddle motion at the same gigahertz frequency and effectively amplifies it, due to its large size.  
\begin{figure}[t] 
\epsfxsize=7.62 cm 
\epsfysize=6.1 cm 
\epsfbox{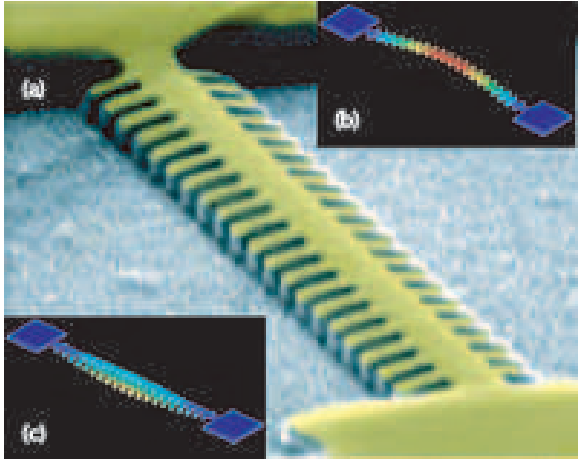} 
\caption{
a) SEM micrograph of the suspended antenna oscillator.  It consists of a central Si beam, $10.7~\mu m$ long and 400 nm wide, as well as two arrays of 500 nm long and 200 nm wide paddles on both sides.  The total thickness of the structure is 245 nm, comprised of the device layer of silicon (185 nm) and the thermally evaporated gold electrode (60 nm), colorized in yellow.  b) Modal simulation of the antenna structure, showing the low frequency ($f \simeq 10 \mbox{ MHz}$) fundamental resonance mode.  c) In the high order collective mode, the paddles vibrate at their own natural frequency ($f\ge 1 \mbox{ GHz}$), and the induced in-phase strain drives the central beam, which acts as an amplifier of the paddle motion.
}  
\end{figure}  

We have fabricated four identical antenna structures from single crystal silicon with e-beam lithography and nanomachining.  When placed at the center of a 16 Tesla superconducting magnet in a dilution fridge, each structure is driven magnetomotively by a Lorentz force $F_{dr}(\omega)= I(\omega)LB$ with current $I(\omega)$ through the electrode of length $L$ in perpendicular magnetic field $B$.  The resulting motion is given by the harmonic oscillator response function  $x(\omega)=F_{dr}(\omega)/[(\omega_0^2-\omega^2+i\omega_0\omega/Q)m]$.  The beam displacement $x$ in magnetic field $B$ induces a voltage $V_{emf}(\omega)=i\xi LB\omega_0x(\omega)$ in the gold electrode. Here $\xi$ is a mode-dependent integration constant \cite{cleland-nature}. We measure $V_{emf}$ using an RF network analyzer.  Combining the expressions for $F_{dr}(\omega)$, $x(\omega)$ and the harmonic oscillator equation, the magnetomotive expression is:
\begin{equation}
V_{emf}(\omega) = {i\omega\xi L^2 B^2 \over {\omega_0^2-\omega^2 + i\omega\omega_0/Q}}I_{dr} (\omega)    
\end{equation} 
\begin{figure}[t] 
\epsfxsize=8.7 cm 
\epsfysize=8 cm 
\epsfbox{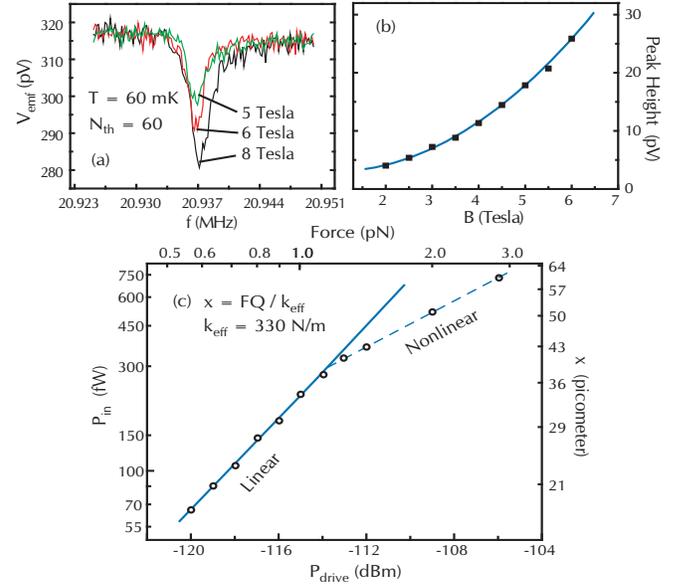} 
\caption{
Low order resonance mode of the antenna structure at 21 MHz with $Q = 11000$.  a)  The induced voltage $V_{emf}$ is an inverted Lorentzian peak on top of a noise background.  Inversion results from extra phase in the coaxial cable impedance.  Three representative sweeps show the resonance peak at 5, 6, and 8 Tesla.  b) As expected from the magnetomotive scheme, Eq. 1, the peak height varies as $B^2$ for constant $I_{dr}$.  c) Plot of the linear power dependence of the resonance peak at 2.5 Tesla field, on a log-log scale.  The corresponding displacement $x$ is linear with driving force $F_{dr}$ for forces below 1.3 pN.  Nonlinearity ensues for larger $F_{dr}$.  The effective spring constant $k_{eff} = m\omega_0^2$ measured on resonance $\omega \rightarrow \omega_0$, is 330 N/m. 
}  
\end{figure}  

A typical low order resonance mode of the antenna structure is measured at 21 MHz, as shown in Fig.~2a.  With the oscillator at the equilibrium temperature of 60 mK, the thermal occupation is $N_{th}\simeq 60$, corresponding to the classical regime.  We verify the expected $B^2$ dependence of the response on the magnetic field $B$ in Fig.~2b, which follows from the magnetomotive relation in Eq.1.  For small forces, the center-beam displacement $x$ on resonance varies linearly with the driving force $F_{dr}$ in accordance with $x(\omega_0)=Q F_{dr}(\omega_0)/k_{eff}$ (Fig.~2c).  The linear fit yields an effective spring constant $k_{eff} = 330\mbox{ N/m}$.  A detailed analysis of all the other complex modes and their shapes as well as their dependence on structure geometry will be given elsewhere.

Thermal equilibrium of the oscillator with the mixing chamber of the dilution cryostat occurs primarily through thermal phonon exchange between the single-crystal oscillator and the bulk and by coupling to the gold electrode on top of the beam through the Kapitza resistance \cite{wellstood}.  The thermal phonon wavelength is given by $\lambda_{th}=\hbar v_s/k_BT$, where $v_s = 5000\mbox{ m/s}$ is the sound velocity in silicon. Free thermal phonon propagation through the silicon beam (width $w =\mbox{ 400 nm}$) can occur down to $T = \hbar v_s/k_Bw = 95\mbox{ mK}$. Thermalization of the oscillator occurs also through the silicon-gold interface. The thermal Johnson noise signal size $S_x^{th}=\sqrt{4k_BTQ \over m_{eff}\omega^3}$ decreases dramatically with increasing frequency. For example, a 1-$\mu$m long silicon beam with a resonance frequency of 1.5 GHz has $S_x^{th}\simeq 3 \times 10^{-18} m/\sqrt{Hz}$ at a temperature of 100 mK, three orders of magnitude smaller than the best sensitivity of $\sim 10^{-15} m/\sqrt{Hz}$ reported to date \cite{cleland-nature,schwab}.

We observe a high-frequency collective mode at 1.48 GHz with $Q \simeq 150$, as shown in Fig.~3a; this is the highest nanomechanical resonance frequency reported to date \cite{huang}.  For temperatures below 100 mK, the occupation factor is $N_{th}\rightarrow 1$, and we expect the onset of non-classical behavior.  However, it is imperative to verify that the mode is classical for larger values of $N_{th}$.  Magnetic field and power sweeps $T = 1000\mbox{ mK}$, which corresponds to $N_{th} \simeq 14$, indeed demonstrate the same $B^2$ field and linear force dependence, as shown Figs.~3b and 3c.  Furthermore, the experimentally determined effective spring constant $k_{eff} = 188\mbox{ N/m}$ is an order of magnitude lower than the estimated value for a 1-GHz straight silicon beam, giving higher displacements as expected from our structure design.

\begin{figure}[t] 
\epsfxsize=8.7 cm 
\epsfysize=8.0 cm 
\epsfbox{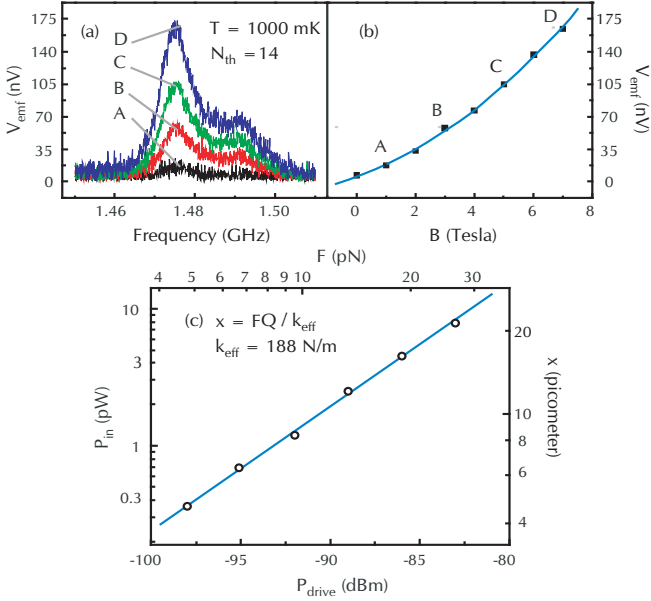} 
\caption{
Classical response of the antenna structure in the 1.48 GHz collective high order mode with $Q = 150$.  At $T = 1000\mbox{ mK}$ the occupation factor is $N_{th} \simeq 14$ for this mode.  a) Four plots (A,B,C, and D) of the induced voltage $V_{emf}$ show the main peak at 1.479 GHz and a smaller secondary  peak at 1.49 GHz, for various values of the magnetic field $B$.  The data is normalized to the noise background at 0 Tesla.  b) In agreement with the magnetomotive expression, the plot of the peak height versus magnetic field $B$ shows quadratic dependence, entirely analogous to the 21 MHz plot.  c) Linear dependence of the induced power $P_{in}$ on driving power $P_{dr}$ shows direct evidence of classical Hooke's law oscillator behavior.  The effective spring constant is $k_{eff} = 188\mbox{ N/m}$.
}  
\end{figure}  

In stark contrast to the classical behavior at $T = 1000\mbox{ mK}$, the collective mode response exhibits non-monotonic magnetic field dependence when the temperature is lowered to $T = 110\mbox{ mK}$, corresponding to $N_{th}\simeq 1$.  In Figs.~4b and 4c, the plots show discrete peak voltage transitions at three values of the magnetic field.  While the transitions do not always occur at exactly the same field values, the size of the jumps remains unchanged, suggesting that the oscillator switches between two well-defined states.  The transition is reproducible for field sweeps in either direction.  Furthermore, in preliminary measurements we have observed spontaneous decay from the upper to the lower state in time, shown in Fig.~4d, with a more detailed investigation forthcoming.  

\begin{figure}[t] 
\epsfxsize=8.7 cm 
\epsfysize=5.8 cm 
\epsfbox{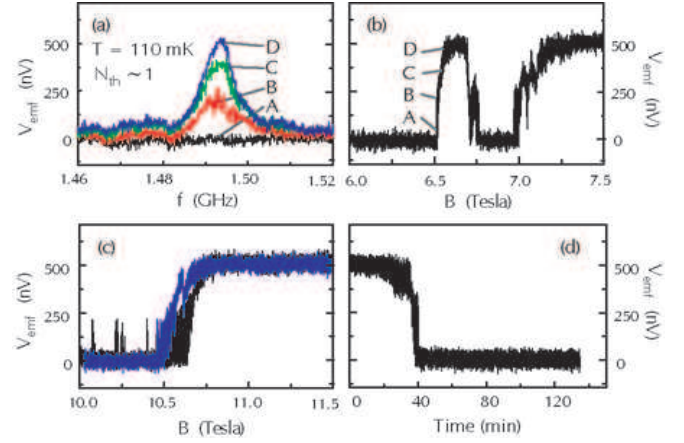} 
\caption{
Non-classical response in the 1.49 GHz collective mode at $T = 110\mbox{ mK}$.  a) Four discrete plots, showing rapid peak growth within the narrow $B$ field range from 6.0 to 6.87 Tesla, which corresponds to 45 to 52 pN of force.  We observe a 0.6 percent frequency blue shift with decreasing temperature.  b) The peak growth is non-monotonic, as seen in this separate continuous field sweep at the center frequency $f_0 = 1.49\mbox{ GHz}$.  The peak undergoes several transitions with amplitude 500 nV between two well-defined states.  The labels A,B,C, and D indicate the approximate location of peak growth in plot a).  c) Two sweeps of the magnetic field, up (black) and down (blue), reproducing the same transition, with occasional transitions to a possible intermediate state.  d) With all parameters held constant, the system spontaneously decays in time, with no further changes observed within a measurement period of 2 hours.
}  
\end{figure}  

The oscillator response in the 1.48-GHz mode in Figs.~3 and 4 demonstrates continuous dependence on
drive force (energy) at 1000 mK and discrete response at 110 mK, corresponding to the classical
regime ($N_{th} \sim 14$) and the quantum regime ($N_{th} \sim 1$) respectively. Since our
experiment involves the measurement of the absolute value of the displacement $|x|$, it is possible
that the discrete transitions show signatures of energy quantization as $|x| \propto \sqrt{E}$, where
$E$ is the oscillator energy. The two distinct saturated states (A and D in Fig.~4b) could then
correspond to the ground state $|0\rangle$ and the first excited state $|1\rangle$ in the energy eigenbasis, respectively. The intermediate jumps between these two states (in the range of
6.5-7 tesla in Fig.~4b) could represent transitions induced by thermal fluctuations, as the temperature
even at $N_{th} \sim 1$ is high enough that the thermal energy $k_BT$ smears the gap $hf$ in the energy levels. If this is true, then future experiments at higher frequencies and lower temperatures, corresponding to a lower $N_{th}$, will show clear transitions from $|0\rangle \rightarrow |1\rangle $. Further reduction in $N_{th}$ could even show multiple transitions $|0\rangle \rightarrow |1\rangle \rightarrow |2\rangle \cdots $, corresponding to access to the first and higer-order excited states.

Another possibility is that the oscillator does not start from the energy eigenbasis, as there is
no {\it a priori} reason for it to be in this preferred basis. As the driving energy is increased
(by increasing the field), the oscillator goes from a certain ground state (point A in Fig. 4b) to a linear combination of $|0\rangle$ and $|1\rangle $ with coefficients corresponding to maximum amplitude of displacement. This process should necessarily involve only the first couple of low-lying states. In fact, for $N_{th} \sim 1$, only the low-lying states ($|0\rangle$ and $|1\rangle $) are substantially populated.

Similar discrete behavior can result classically due to transitions between two nonlinear bistable beam states. After an extensive study of bistability in nanobeams\cite{badzey}, we rule out this mechanism in this case because the response of the structure at the drive level used in the experiment is manifestly linear, seen from the Lorentzian peak shape (Fig.~4a). However, there may be other semiclassical mechanisms which involve both nonlinearity and quantum effects \cite{thorwart}.  

In order to clearly establish whether or not the observed discrete features are quantum mechanical in origin, a complete theory is needed.  For example, simulation of the mechanics of the antenna structure containing fifty billion silicon atoms must involve multiscale modeling, including molecular dynamics at the atomic level and finite element analysis in the continuum elastic theory.  Understanding the measurement of the quantum system itself requires identifying the components of the quantum measurement process: quantum system, measuring apparatus, and their interaction.  Coupling to the environement, which introduces decoherence and dissipation, must be included in the theoretical analysis.  On the experimental side, an irrefutable observation of quantum mechanical transitions will require measurements at higher frequencies and lower temperatures ($N_{th} < 1$) with a clear gap between the energy levels. Detailed study of the noise in the discrete states will also help elucidate the possible quantum mechanical origin of the oscillator response.  
 
In conclusion, we have measured the displacement in a series of nanomechanical oscillators with 1.4-GHz resonance frequencies down to 110 mK temperatures. While the low frequency mode at 21 MHz shows classical behavior with expected drive dependence ($\propto B^2$), the 1.49 GHz mode displays non-monotonic dependence on driving force at a temperature which corresponds to a thermal occupation number $N_{th} \rightarrow 1$. Our experimental data indicates the first observation of quantum displacement in macroscopic nanomechanical oscillators. Our work is supported by the NSF, ARL, ACS and the Sloan Foundation.


\end{document}